\documentclass[twocolumn,aps,prc,showpacs,superscriptaddress,preprintnumbers,floatfix,nofootinbib,longbibliography]{revtex4-1}
\usepackage{multirow}
\usepackage{hhline}
\usepackage{bbm}
\usepackage{epsfig}
\usepackage{graphics}  
\usepackage{graphicx}  
\usepackage{dcolumn}   
\usepackage{bm}        
\usepackage{amssymb}   
\usepackage{amsmath}   
\usepackage{amsfonts}
\usepackage[percent]{overpic}
\usepackage{natbib}
\usepackage{hyperref}
\usepackage[usenames]{color}

\newcommand{\pT}{p_\mathrm{T}}
\newcommand{\mpT}{\langle p_\mathrm{T}\rangle}
\newcommand{\dmpT}{\langle\langle p_\mathrm{T} \rangle\rangle}

\newcommand{\mnpart}{\langle \mathrm{N}_\mathrm{part} \rangle}

\newcommand{\auau}{Au+Au }

\usepackage{tikz,xcolor,hyperref}
\definecolor{lime}{HTML}{A6CE39}
\DeclareRobustCommand{\orcidicon}{
	\begin{tikzpicture}
	\draw[lime, fill=lime] (0,0) 
	circle [radius=0.16] 
	node[white] {{\fontfamily{qag}\selectfont \tiny ID}};
	\draw[white, fill=white] (-0.0625,0.095) 
	circle [radius=0.007];
	\end{tikzpicture}
	\hspace{-2mm}
}
\foreach \x in {A, ..., Z}{%
	\expandafter\xdef\csname orcid\x\endcsname{\noexpand\href{https://orcid.org/\csname orcidauthor\x\endcsname}{\noexpand\orcidicon}}
}
\foreach \x in {A, ..., Z}{%
	\expandafter\xdef\csname orcid\x\endcsname{\noexpand\href{https://orcid.org/\csname orcidauthor\x\endcsname}{\noexpand\orcidicon}}
}

\begin{document}

\title{Energy dependence of transverse momentum fluctuations in Au+Au collisions from a multiphase transport model}

\newcommand{\hinst}{Institute of Nuclear Science and Technology, Henan Academy of Sciences, Zhengzhou, 450015, China}
\newcommand{\moe}{Key Laboratory of Nuclear Physics and Ion-beam Application (MOE), and Institute of Modern Physics, Fudan University, Shanghai 200433, China}
\newcommand{\fudan}{Shanghai Research Center for Theoretical Nuclear Physics, NSFC and Fudan University, Shanghai 200438, China}
\newcommand{\sbu}{Department of Chemistry, Stony Brook University, Stony Brook, NY 11794, USA}
\author{Liuyao Zhang\orcidA{}}\affiliation{\hinst}\affiliation{\moe}\affiliation{\fudan}
\author{Jinhui Chen\orcidB{}}\email{chenjinhui@fudan.edu.cn}\affiliation{\moe}\affiliation{\fudan}
\author{Chunjian Zhang\orcidC{}}\email{chunjianzhang@fudan.edu.cn}\affiliation{\moe}\affiliation{\fudan}\affiliation{\sbu}

\begin{abstract}
Event-by-event mean transverse momentum fluctuations ($\mpT$) serve as a sensitive probe of initial state overlap geometry and energy density fluctuations in relativistic heavy-ion collisions. We present a systematic investigation of $\mpT$ fluctuations in \auau collisions at $\mathrm{\sqrt{s_{NN}}}$ = 3.019.6 GeV, examining their centrality and energy dependence with the framework of an improved multiphase transport (AMPT) model. The centrality dependence of the $\pT$ cumulants up to fourth order deviates significantly from simple powering-law scaling. 
Scaled cumulants are performed, with variances aligning well with the trends observed in the experimental data. Employing a two-subevent method, short-range correlations are slightly suppressed compared to the standard approach. Furthermore, baryons exhibit more pronounced $\mpT$ fluctuations than mesons, potentially attributable to the effect of radial flow. These results provide referenced insights into the role of initial state fluctuations across different energies in heavy-ion collisions.

\end{abstract}
\maketitle

\section{Introduction} 
Event-by-event (EbE) fluctuations have been proposed as a useful probe for exploring initial state and thermalization in heavy-ion collision at relativistic energies~\cite{NA49:1999inh,PHENIX:2003ccl, 
STAR:2003cbv,ALICE:2023tej,Gavin:2003cb, Bozek:2012fw,Zhang:2021vvp}, offering critical insights into the properties of Quark-Gluon Plasma (QGP) formation~\cite{Heiselberg:2000fk}. These fluctuations are particularly significant for exploring the Quantum Chromodynamics (QCD) phase diagram, including the QGP-to-hadron gas transition and the potential existence of a critical point in strongly interacting matter~\cite{STAR:2010vob,Chen:2024aom}. Fluctuations in the positions of participating nucleons lead to EbE fluctuations in the initial state, which propagate to final-state observables during system expansion~\cite{Alver:2010gr, Teaney:2010vd,Schenke:2020uqq,Jia:2021tzt}. The temperature fluctuations associated with phase transition in the QCD phase diagram can manifest themselves in event-wise mean transverse momentum ($\mpT$) fluctuations of the final-state particles~\cite{Stodolsky:1995ds}. $\mpT$ is sensitive to the initial energy density and inversely proportional to the size of the overlap region~\cite{Shuryak:1997yj,Jia:2021qyu,Samanta:2023amp}, while it is also influenced by collective behavior, fluctuations in the participant nucleons, and other final-state effects~\cite{Bhatta:2021qfk}. 

The dynamical $\mpT$ fluctuations estimated by variance have been extensively studied in the Super Proton Synchrotron (SPS)~\cite{NA49:1999inh,
CERES:2003sap,NA49:2003hxt,CERES:2008wlj, NA49:2008fag}, Relativistic Heavy-Ion Collider (RHIC)~\cite{PHENIX:2002aqz,
PHENIX:2003ccl, STAR:2003cbv, STAR:2005vxr, STAR:2005xhg}, and the Large Hadron Collider (LHC)~\cite{ALICE:2014gvd,ALICE:2023tej,ALICE:2024apz}. 
These results reveal a universal multiplicity dependence, where correlations are progressively diluted with increasing participant numbers, consistent with dominance by particle pairs originating from the same nucleon-nucleon collisions. Scaled dynamical correlations exhibit minimal dependence on the collision energy, as observed in both small and large collisions~\cite{ALICE:2014gvd,STAR:2013sov}.
Recently, higher-order $\mpT$ fluctuations have also been investigated in experiments to uncover the intricate mechanisms underlying the observed fluctuations. The ALICE experiment reports a positive skewness in $\mpT$ fluctuations in minimum-bias $pp$ collisions and across all centralities in large collision systems~\cite{ALICE:2023tej}, consistent with hydrodynamic simulation predictions~\cite{Gale:2013da, Bhatta:2021qfk, Giacalone:2020lbm}. Furthermore, the kurtosis of $\mpT$ fluctuations in most central Pb+Pb collisions aligns with that of Gaussian distribution, suggesting the formation of a locally thermalized system~\cite{ALICE:2023tej}. The ATLAS collaboration has also observed distinct changes in the mean, variance, and skewness of $\mpT$ distributions in ultra-central Pb+Pb and Xe+Xe collisions, demonstrating a clear disentanglement of geometrical and intrinsic components through the analysis of the speed of sound~\cite{ATLAS:2024jvf}. Nevertheless, such EbE $\mpT$ fluctuations have also been proposed as a tool to study nuclear structure effect~\cite{STAR:2024wgy,Zhang:2022sgk,Jia:2022ozr,Jia:2021qyu,Xu:2022ikx}.

From a theoretical perspective, $\mpT$ fluctuations have been explored within the color glass
condensate (CGC) framework~\cite{Gelis:2010nm,Weigert:2005us},
which has demonstrated strong concordance with data in semicentral and central collisions~\cite{Gavin:2011gr}. The UrQMD model~\cite{Bleicher:1999xi,Bass:1998ca}, incorporating only hadronic transport, successfully reproduces the relative dynamical correlation in Au+Au central collisions at both RHIC~\cite{STAR:2019dow} and LHC~\cite{ALICE:2014gvd} energies. 
Both PYTHIA8~\cite{Skands:2014pea} and EPOSLHC~\cite{Werner:2013tya} also have successfully modeled relative dynamical correlation at the LHC energy, and provided insight into the interpretation of the data~\cite{ALICE:2024apz}. Similarly, the AMPT model, particularly with the string-melting version demonstrates good agreement with experimental measurements at the LHC energy~\cite{ALICE:2014gvd, ALICE:2024apz, Xu:2020pxj}. In higher-order $\mpT$ fluctuations, HIJING simulations exhibit a power-law scaling on charged particle multiplicity in both skewness and kurtosis, consistent with a superposition of independent sources~\cite{Bhatta:2021qfk}. Moreover, hydrodynamic calculations can explain the measurements of both skewness and kurtosis quantitively from semicentral to central collisions at LHC energy~\cite{ALICE:2023tej,Giacalone:2020lbm}.

Given that dynamical fluctuations in two-particle $\pT$ correlations have been measured during the RHIC Beam Energy Scan (BES) program~\cite{STAR:2019dow}, which aims to investigate the QCD first-order phase transition and identify the possible critical point, there remains a notable absence of dynamic studies focused on higher-order $\mpT$ fluctuations stemming from model calculation.

In this paper, we study the collision energy dependence of $\mpT$ fluctuations, from mean to kurtosis, using an improved AMPT model. Variances are calculated using both the standard and two-subevent methods, revealing underlying decorrelation effects in $\mpT$ fluctuations. We further investigate baryon and meson fluctuations across different collision energies and systematically examine the effects of various configurations and acceptances. The paper is organized as follows. Section~\ref{sec:Methodology} introduces the formula for calculating $\pT$ cumulants using standard and subevent methods. Sec.~\ref{sec:ampt} describes the improvements to the AMPT model, incorporating system size and collision energy dependences in the Lund fragmentation parameter. The main results are presented in Sec.~\ref{sec:results}, followed by a summary in Sec.~\ref{sec:discussion}. 

\section{Methodology}
\label{sec:Methodology} 
Following the $\mpT$ cumulants approach in Refs.~\cite{Bhatta:2021qfk,Jia:2017hbm}, the $n$-particle $\pT$ correlator in one event is defined as
\begin{equation}
c_n =\frac{\sum\limits_{i_1 \neq \cdots \neq i_n} w_{i_1} \cdots w_{i_n}\left(p_{\mathrm{T}, \mathrm{i}_1}-\left\langle\left\langle p_{\mathrm{T}}\right\rangle\right\rangle\right) \cdots\left(p_{\mathrm{T}, \mathrm{i}_{\mathrm{n}}}-\left\langle\left\langle p_{\mathrm{T}}\right\rangle\right\rangle\right)}{\sum\limits_{i_1 \neq \cdots \neq i_n} w_{i_1} \cdots w_{i_n}}.
\label{eq:ck_define}
\end{equation}
Here, $w_{i}$ represents the weight of particle $i$, and the relation expands algebraically into a simple polynomial form, 
\begin{equation}
\begin{aligned}
&p_{m k}  = \sum\limits_i w_i^k p_i^m / \sum\limits_i w_i^k, \quad \tau_k=\frac{\sum\limits_i w_i^{k+1}}{\left(\sum\limits_i w_i\right)^{k+1}}, \\
&\bar{p}_{1k} \equiv  p_{1 k}-\left\langle\left\langle p_{\mathrm{T}}\right\rangle\right\rangle, \\
&\bar{p}_{2k} \equiv  p_{2 k}-2 p_{1 k}\left\langle\left\langle p_{\mathrm{T}}\right\rangle\right\rangle+\left\langle\left\langle p_{\mathrm{T}}\right\rangle\right\rangle^2, \\
&\bar{p}_{3k} \equiv  p_{3 k}-3 p_{2 k}\left\langle\left\langle p_{\mathrm{T}}\right\rangle\right\rangle+3 p_{1 k}\left\langle\left\langle p_{\mathrm{T}}\right\rangle\right\rangle^2-\left\langle\left\langle p_{\mathrm{T}}\right\rangle\right\rangle^3, \\
&\bar{p}_{4k} \equiv  p_{4 k}-4 p_{3 k}\left\langle\left\langle p_{\mathrm{T}}\right\rangle\right\rangle+6 p_{2 k}\left\langle\left\langle p_{\mathrm{T}}\right\rangle\right\rangle^2 
-4 p_{1 k}\left\langle\left\langle p_{\mathrm{T}}\right\rangle\right\rangle^3 \\
&  + \left\langle\left\langle p_{\mathrm{T}}\right\rangle\right\rangle^4.
\end{aligned}
\end{equation}
Where $\dmpT =\bar{p}_\mathrm{11}$ is the $\mpT$ averaged over the event ensemble, with $p \equiv \pT$. 

Using auxiliary variables, the correlator in Eq.~\ref{eq:ck_define} can be expressed as,
\begin{equation}
\begin{aligned}
&c_2 =\frac{\bar{p}_{11}^2-\tau_1 \bar{p}_{22}}{1-\tau_1}, \\
&c_3 =\frac{\bar{p}_{11}^3-3 \tau_1 \bar{p}_{22} \bar{p}_{11}+2 \tau_2 \bar{p}_{33}}{1-3 \tau_1+2 \tau_2}, \\
&c_4 =\frac{\bar{p}_{11}^4-6 \tau_1 \bar{p}_{22} \bar{p}_{11}^2+3 \tau_1^2 \bar{p}_{22}^2+8 \tau_2 \bar{p}_{33} \bar{p}_{11}-6 \tau_3 \bar{p}_{44}}{1-6 \tau_1+3 \tau_1^2+8 \tau_2-6 \tau_3}.
\end{aligned}
\end{equation}
where particles are selected from pseudorapidity range $|\eta| <$ 1.0 and 0.5 $ \leq p_\mathrm{T} \leq$ 3.0 GeV/c, considering all unique combinations within each event. The advantage of employing multiparticle $\pT$ correlator~\cite{Voloshin:2001ei, Voloshin:2002ku} is that they yield zero values for events with randomly sampled particles, thereby effectively isolating the non-statistical fluctuations of interest.

In addition, two-subevent method (2sub) is employed, where particle combinations are selected from two $\eta$-separated subevents, a (-1.0 $<\eta<\eta_1$) and c ( $\eta_2<\eta<$ 1.0). 
This $\eta$ gap could reduce short-range correlations. To evaluate the impact of these correlations on high-order cumulants, the gap is varied to 0.4. The resulting correlations are given by
\begin{equation}
\begin{aligned}
&c_{2, \text {sub}} =\left(\bar{p}_{11}\right)_a\left(\bar{p}_{11}\right)_c \\
\end{aligned}
\label{eq:cn_sub}
\end{equation}
where the subscripts in Eq.~\ref{eq:cn_sub} indicate the subevents from which particles are selected. Since there are two alternative ways to calculate two-subevent $c_{3}$, the final value is taken as the average, 
$c_{3,2 \mathrm{sub}}=(c_{3,2 \mathrm{sub1}}+c_{3,2 \mathrm{sub2}})/2$. 

In Eq.~\ref{eq:ck_define}, the $n$-particle $p_\mathrm{T}$ correlator varies with incident energy and collision centrality. To account for these variations, the measured correlations for variance, 
skewness, and kurtosis are scaled by $\dmpT$, respectively, yielding dimensionless scaled cumulants.
\begin{equation}
\begin{aligned}
&k_2 =\frac{\sqrt{\left\langle c_2\right\rangle}}{\left\langle\left\langle p_{\mathrm{T}}\right\rangle\right\rangle}, \\ 
&k_3 =\frac{\sqrt[3]{\left\langle c_3\right\rangle}}{\left\langle\left\langle p_{\mathrm{T}}\right\rangle\right\rangle}, \\
&k_4 =\frac{\sqrt[4]{\left\langle c_4\right\rangle-3\left\langle c_2\right\rangle^2}}{\left\langle\left\langle p_{\mathrm{T}}\right\rangle\right\rangle}.
\end{aligned}
\end{equation}

For the subevent method, the corresponding scaled cumulants are given by
\begin{equation}
\begin{aligned}
&k_{2,2 \text {sub}} =\sqrt{\frac{\left\langle c_{2, 2\text {sub}}\right\rangle}{\left\langle\left\langle p_{\mathrm{T}}\right\rangle\right\rangle_a\left\langle\left\langle p_{\mathrm{T}}\right\rangle\right\rangle_c}}. \\
\end{aligned}
\end{equation}

\section{AMPT model}
\label{sec:ampt}
\begin{figure*}[htb]
\centering
\includegraphics[scale=0.6]{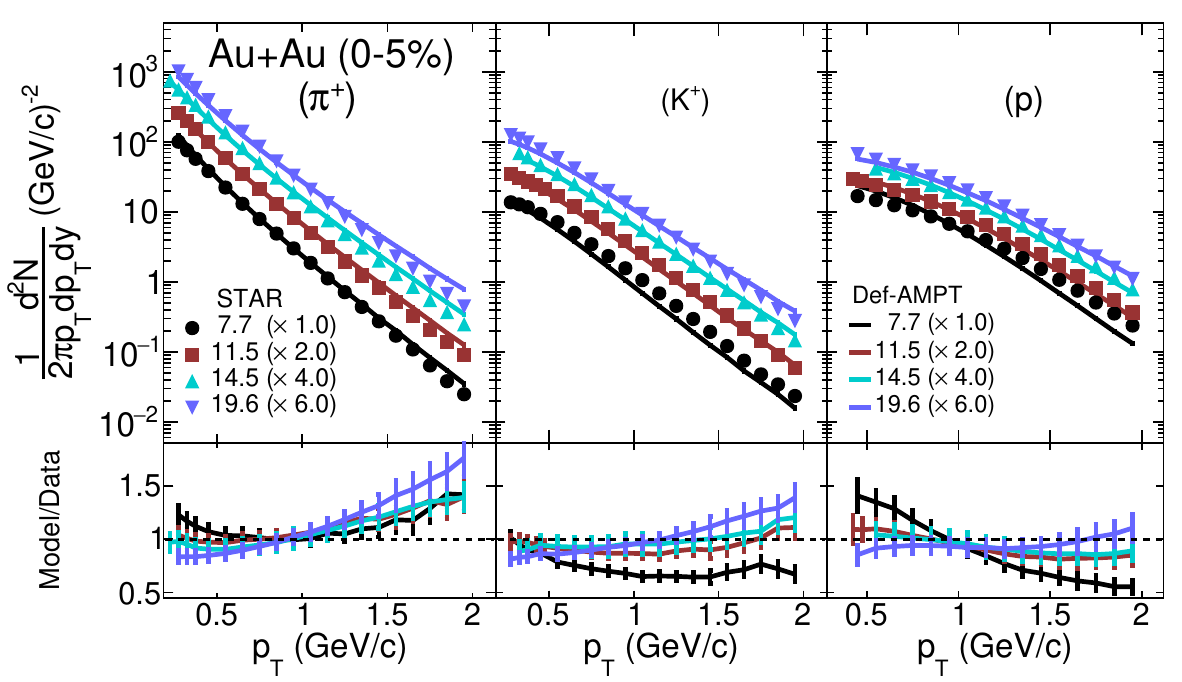}
\caption{(Color online) Transverse momentum spectra at midrapidity ($|y|<$ 0.1) for $\pi^{+}$ (left), $K^{+}$ (middle), and protons (right) in Au+Au collisions at $\mathrm{\sqrt{s_{NN}} =}$ 7.7, 11.5, 14.5, and 19.6 GeV for the 0–5\% centrality class. The spectra are scaled for clarity, with data from Ref.~\cite{STAR:2017sal}. The lower panels show the model to data ratio.}
\label{fig:pTdis}
\end{figure*}
AMPT is an effective Monte Carlo framework, which is extensively used to study relativistic heavy-ion collisions at SPS, RHIC and the LHC energies~\cite{Zhang:1999bd,Lin:2004en,Zhang:2018ucx,Zhang:2019bkf,Zhang:2021ygs,Zhang:2021kxj,Giacalone:2021udy,Jia:2022qgl,Mendenhall:2021maf,Zhang:2022fum,Zheng:2024xyv,Wang:2024vjf,Zhang:2024vkh,Pu:2023xrl}. The AMPT model has two versions: the default version (denoted as Def-AMPT) and the string-melting (denoted as SM-AMPT) version. Both versions comprise four main dynamic components: (i) a fluctuating initial condition from the Heavy
Ion Jet INteraction Generator (HIJING) model~\cite{Wang:1991hta}, (ii) parton cascade simulations using the Zhang's Parton cascade (ZPC) model~\cite{Zhang:1997ej}, which includes only 2 to 2 elastic parton processes, (iii) hadronization via the Lund string fragmentation or a quark coalescence model, 
and (iv) hadron scattering described by A Relativistic 
Transport (ART) model~\cite{Li:1995pra}. 
This study employs the latest version, the AMPT-v1.26t9b-v2.26t9b~\cite{Lin:2021mdn,Magdy:2021ocp}, 
with the QCD coupling constant $\alpha_{s}$ = 0.33, the screening mass $\mu = $2.265 $\text{fm}^{-1}$ corresponding to a parton scattering cross section of 3.0 mb in the ZPC.

The default AMPT model is utilized primarily in this study because of its success in describing the transverse momentum spectra 
of identified particles in heavy-ion collisions at SPS and RHIC energies~\cite{Shao:2020sqr, Lin:2021mdn}. The $\dmpT$ is expected to increase with centrality due to higher initial temperatures in more central collisions~\cite{STAR:2017sal}.
However, neither the original default version nor the string-melting version of AMPT adequately reproduces the $\dmpT$ as a function of centrality near midrapidity~\cite{Ma:2016fve,Zhang:2021vvp}. To address the reversed trend predicted by AMPT, the Lund fragmentation parameters were tuned to introduce a system-size dependence. According to the Lund string fragmentation mechanism,  
the average squared transverse momentum of produced particles is proportional to the
string tension $\kappa$, which reflects the energy stored per unit length of the string and is determined by the fragmentation parameters $a_{L}$ and $b_{L}$~\cite{Lin:2004en}, as shown in Eq.~\ref{eq:meanpT}.
\begin{equation}
\begin{aligned}
\kappa \propto \left\langle p_{\mathrm{T}}^2\right\rangle=\frac{1}{b_L\left(2+a_L\right)}.
\end{aligned}
\label{eq:meanpT}
\end{equation}

Inspired by Ref.~\cite{Zhang:2021vvp}, the Lund string fragmentation parameter $b_{L}$ is treated as a local variable dependent on the nuclear thickness functions of two nuclei, rather than being a constant parameter. Consequently, $b_{L}$ exhibits an approximate linear dependence on the impact parameter. For simplicity, $b_L$ is tuned in this study to vary linearly with the impact parameter, closely reproducing the linear relationship observed in Ref.~\cite{Zhang:2021vvp}. This tuning enables the AMPT model to reasonably describe the centrality dependence of $\dmpT$. The $\dmpT$ also shows a systematic dependence on incident energy~\cite{STAR:2017sal}, with higher incident energies yielding larger $\dmpT$ due to enhanced collective effects. To account for this, $b_{L}$ is further adjusted to increase slightly with collision energy, allowing the improved AMPT to reproduce the energy dependence of $\dmpT$, as shown in Fig.~\ref{fig:ck} (a).

In this study, $\pT$ cumulants are calculated for $\pi^{\pm}$, $K^{\pm}$ and $p(\bar{p}$) within $|\eta|<$ 1.0 and 0.5 $\leq\pT\leq$ 3.0 GeV/$c$. Centrality classes in Au+Au collisions at $\mathrm{\sqrt{s_{NN}} = 7.7}$$\text{–}$19.6 GeV are defined by the number of participants $N_\mathrm{part}$, directly extracted from AMPT calculation. Events are divided into the following centrality classes: 0–5\%, 5–10\%, 10–20\%, 20–30\%, 30–40\%, 40–50\%, 50–60\%, 60-70\%, and 70–80\%. The mean number of participating nucleons, $\langle N_\mathrm{part} \rangle$, corresponds to these centrality intervals. 

\section{Results and discussions} 
\label{sec:results}

To evaluate the validity of the enhanced AMPT model relative to its original version, we first examine the $\pT$ distributions of select particles. Figure~\ref{fig:pTdis} presents the $\pT$ spectra for $\pi^{+}$ (left), $K^{+}$ (middle), and protons (right) within the midrapidity interval ($|y|<$ 0.1) for Au+Au collisions at $\mathrm{\sqrt{s_{NN}}}=$ 7.7, 11.5, 14.5, and 19.6 GeV, specifically in the 0-5\% centrality class. The ratios of the AMPT results to experimental data are computed to quantify discrepancies. Overall, the improved AMPT model qualitatively reproduces the experimental observables for hadrons across a broad $\pT$ spectrum, spanning a variety of incident energies. For clarity, the $\pT$ spectra for different energies have been appropriately scaled.

\begin{figure}[htb]
\centering
\includegraphics[scale=0.48]{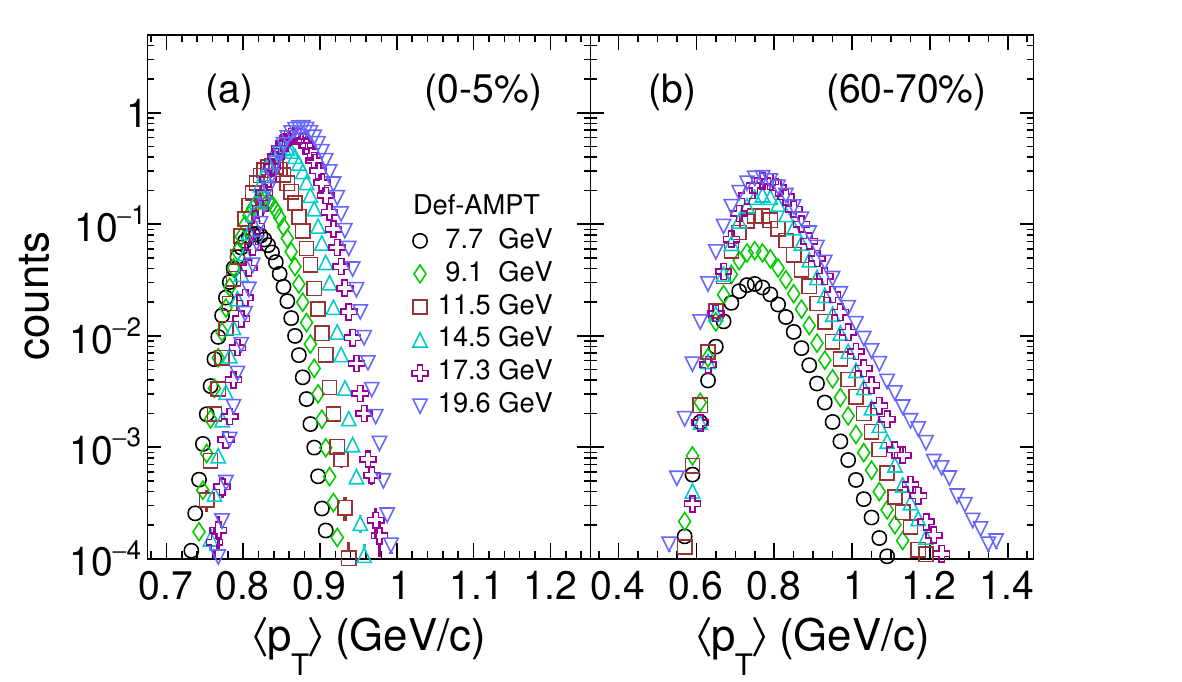}
\caption{(Color online) Distributions of $\langle p_\mathrm{T} \rangle$ from AMPT simulations for Au+Au collisions at $\mathrm{\sqrt{s_{NN}} = }$ 7.7, 9.1, 11.5, 14.5, 17.3 and 19.6 GeV, calculated for $\pi^{\pm}$, $K^{\pm}$, and $p(\bar{p}$) within 0.5 $ \leq p_\mathrm{T} \leq $ 3.0 GeV/$c$ and ($|\eta|<$ 1.0) for the 0–5\% (a), and 60–70\% (b) centrality classes, respectively.}
\label{fig:mpT_ebe}
\end{figure}

The $\mpT$ fluctuations can be straightforwardly studied through its event-wise distributions. Figure~\ref{fig:mpT_ebe} shows the $\mpT$ distributions for the 0–5\% (a) and 50–60\% (b) centrality classes in Au+Au collisions at $\mathrm{\sqrt{s_{NN}}=}$ 7.7, 9.1, 11.5, 14.5, 17.3, and 19.6 GeV. The analysis reveals $\mpT$ values for both centrality classes steadily increase with higher collision energies. In peripheral collisions (50–60\% centrality), the $\mpT$ distributions exhibit greater variances, indicating enhanced fluctuations. Additionally, these distributions show pronounced asymmetry, characterized by a significant rightward tail, suggesting stronger positive skewness in peripheral collisions compared to central collisions. This behavior is consistent with experimental observations at RHIC and at LHC~\cite{PHENIX:2002aqz,Tripathy:2022vwb, ALICE:2013rdo, ALICE:2023tej}.

\begin{figure*}[htb]
\centering
\includegraphics[scale=0.6]{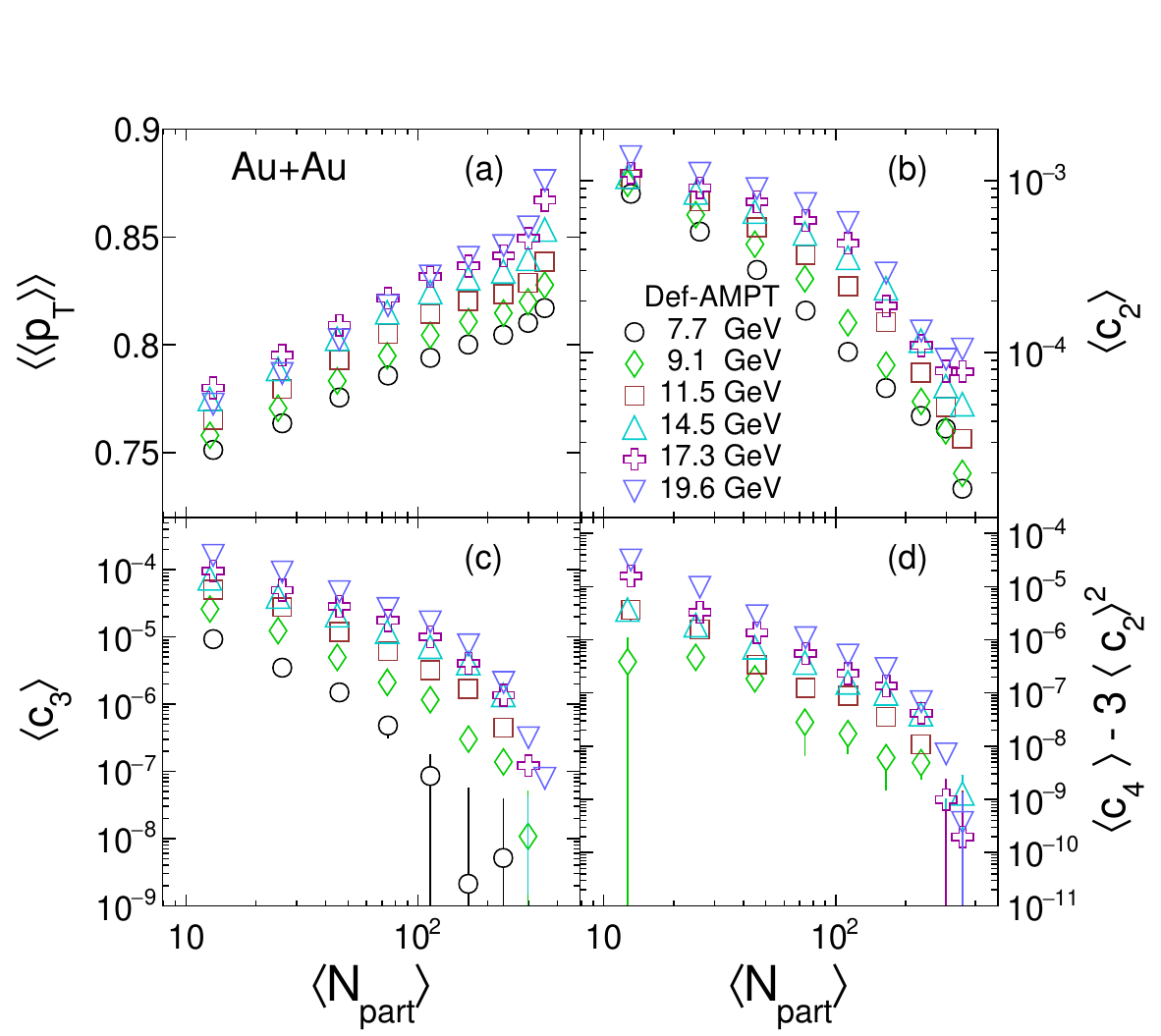}
\caption{(Color online) The $\pT$ cumulants, including event-ensemble transverse momentum $\dmpT$ (a), variance $\langle c_{2} \rangle$ (b), skewness $\langle c_{3} \rangle$ (c), and kurtosis $\left\langle c_4\right\rangle-3\left\langle c_2\right\rangle^2$ (d), are calculated for selected particles within 0.5 $\leq \pT \leq$ 3.0 GeV/c and $|\eta|<$ 1.0 as a function of $\mnpart$ from peripheral to central collisions across difference collision energies in 7.7, 9.1, 11.5, 14.5, 17.3, and 19.6 GeV.}
\label{fig:ck}
\end{figure*}

Figure~\ref{fig:ck} presents the event-ensemble $\pT$ cumulants, including the mean $\dmpT$ (a), variance $\langle c_{2} \rangle$ (b), skewness $\langle c_{3} \rangle$ (c), and kurtosis $\left\langle c_4\right\rangle-3\left\langle c_2\right\rangle^2$ (d) of $\pT$ distributions in Au+Au collisions over beam energies from 7.7 to 19.6 GeV. These cumulants are shown as a function of collision centrality, from peripheral 70–80\% to the most central 0–5\%. The $\dmpT$ values from the enhanced AMPT model exhibit a pronounced centrality dependence, with higher values observed in more central collisions, consistent with previously published AMPT calculations~\cite{Zhang:2021vvp} and the experimental measurements from STAR~\cite{STAR:2017sal}. Additionally, a significant beam energy dependence is observed, where lower collision energies yield smaller $\dmpT$ values across all centrality bins. Similar dependences on centrality or multiplicity, and beam energies for the $\dmpT$ distributions have also been reported at RHIC~\cite{STAR:2017sal} and the LHC energies~\cite{ALICE:2013rdo, ALICE:2024apz}. These trends can be attributed to increased particle production and stronger collective flow in more central collisions, where the overlap region of the colliding nuclei is larger. Higher beam energies deposit more energy in the collision zone, leading to higher temperatures and stronger radial flow in the strongly expanding QGP fireball~\cite{Parida:2024ckk}. 

The second-order $\pT$ cumulant, $\langle c_2 \rangle$, exhibits a non-zero values across 7.7–19.6 GeV, as shown in panel (b) of Fig.~\ref{fig:ck}, suggesting significant EbE fluctuations. An inverse dependence of variance on centrality is consistently observed across all energies, aligning with findings from the STAR and HIJING results~\cite{STAR:2005vxr}. However, the HIJING calculations show an opposite energy dependence in peripheral versus central collisions, in contrast to the consistent energy dependence observed in published STAR measurements. Notably, our findings exhibit a systematic collision energy dependence across various centrality intervals.
The dilution of correlations with increasing centrality may result from a reduction in particle-pair correlations if they are dominated by particles originating from the same nucleon-nucleon collisions. This interpretation is intuitively cross-validated by $\mpT$ distributions across 60–70\% and 0–5\% centrality classes, as shown in Fig.~\ref{fig:mpT_ebe}. 

The higher-order cumulants, skewness $\langle c_{3} \rangle$ and kurtosis $\left\langle c_4\right\rangle-3\left\langle c_2\right\rangle^2$ are presented in panels (c) and (d) of Fig.~\ref{fig:ck}, respectively. Both correlators exhibit significant dependences on centrality and beam energy across $\sqrt{s_{\rm{NN}}}=$ 7.7–19.6 GeV. Their magnitudes decrease by more than one order of magnitude with increasing centrality classes $\mnpart$. Due to the lower statistics, we admit that the kurtosis calculations at 7.7 GeV are not shown.
 
\begin{figure*}[htb]
\centering
\includegraphics[scale=0.78]{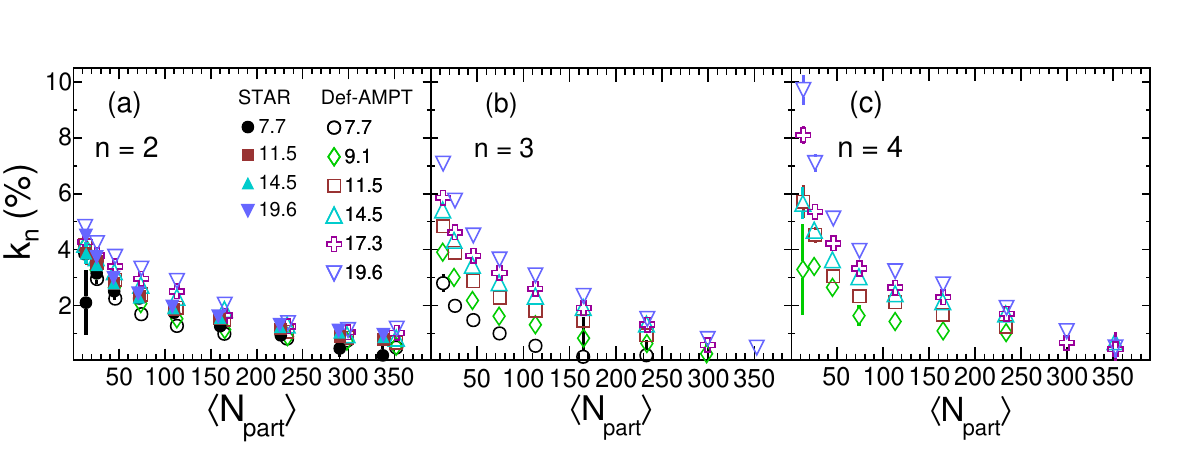}
\caption{(Color online) The scaled variance $k_2$ (a), skewness $k_3$ (b), and kurtosis $k_4$ (c) as a function of centrality in Au+Au collisions across different collision energies with the AMPT model.}
\label{fig:scaledVarSkeKur}
\end{figure*}

To mitigate the influence of variations in $\dmpT$ with incident energy and/or centrality, scaled quantities including variances $k_2$, skewness $k_3$, and kurtosis $k_4$, normalized by $\dmpT$, are shown as functions of centrality for Au+Au collisions across $\sqrt{s_{\rm{NN}}}=$ 7.7-19.6 GeV in Fig.~\ref{fig:scaledVarSkeKur}, respectively. Similarly, the $k_n$ (n=2,3,4) exhibits significant centrality and energy dependence. Notely, the scaled variance $k_2$ qualitatively reproduce the trends observed in the published STAR experiment~\cite{STAR:2019dow}, and also demonstrate remarkable agreement with the available STAR measurements at 7.7, 11.5, and 14.5 GeV, validating the model's predictive capabilities in these energy ranges. However, a systematic overestimation merges, particularly in peripheral to mid-central collisions. Additionally, these scaled variances $k_n$ exhibit an approximate power-law behavior concerning centrality dependence across different collision energies, supporting the independent source pictures.

\begin{figure*}[htb]
\centering
\includegraphics[scale=0.78]{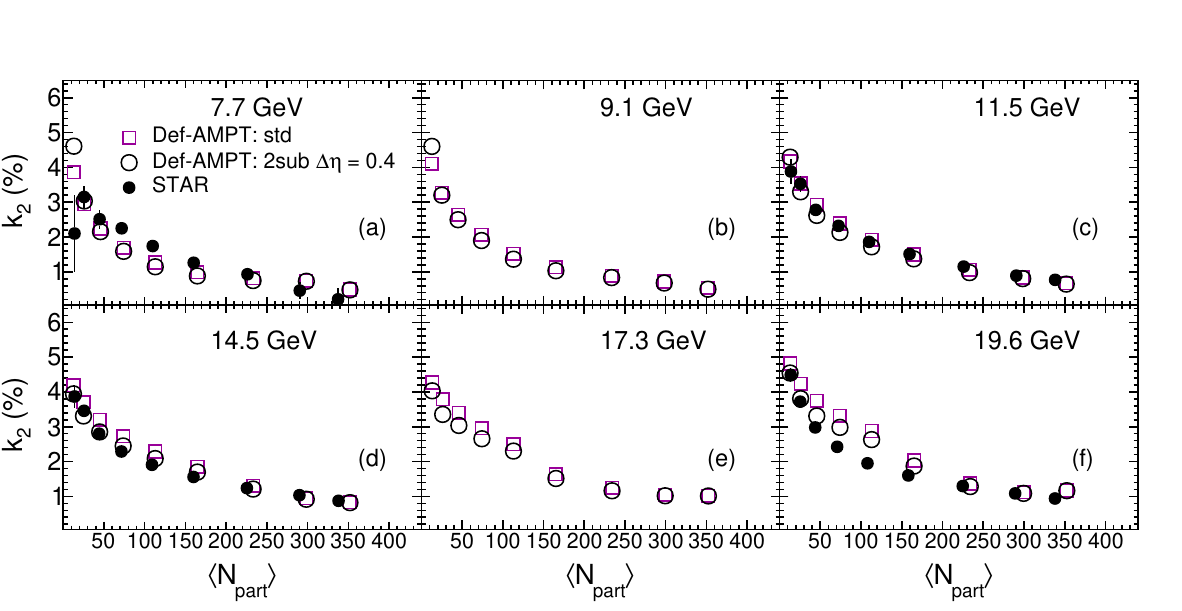}
\caption{(Color online) The scaled variance $k_2$ as a function of centrality $\mnpart$ in Au+Au collisions at $\sqrt{s_\mathrm{NN}} =$ 7.7, 9.1, 11.5, 14.5, 17.3, and 19.6 GeV is calculated within AMPT model using both the standard method (std) and two-subevent (2sub) method, with $\Delta \eta=0.4$.}
\label{fig:scaledVar_2sub}
\end{figure*}

\begin{figure*}[htb]
\centering
\includegraphics[scale=0.78]{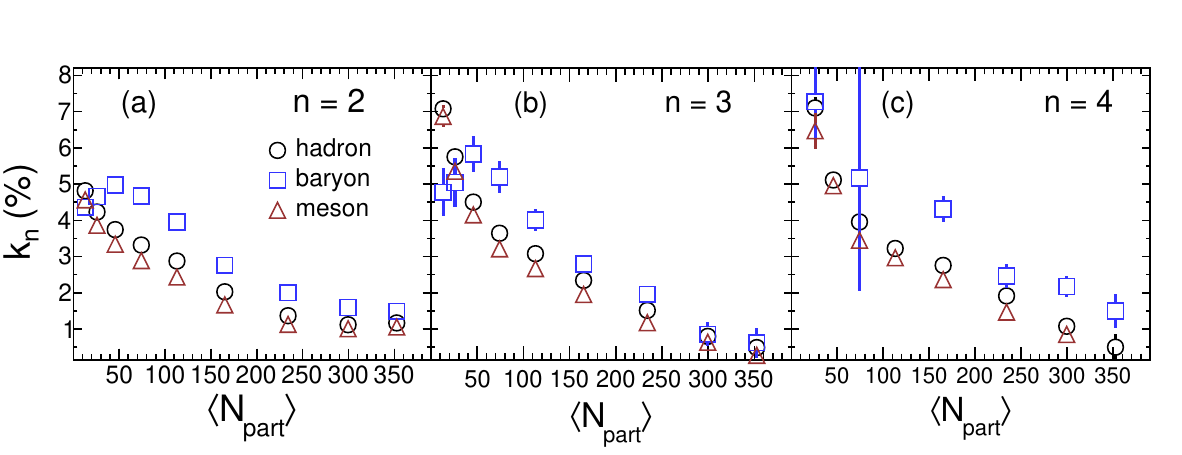}
\caption{(Color online) The scaled variance $\mathrm{k_2}$ (a), skewness $\mathrm{k_3}$ (b) and kurtosis $\mathrm{k_4}$ (c) are shown as a function of centrality $\mnpart$ for ``hadron", ``baryon" and ``meson" at a collision energy of 19.6 GeV.}
\label{fig:scaledVarSkeKur_BM}
\end{figure*}

In order to minimize the impact of short-range correlations from jets and resonance decays, a two-subevent method is employed and 
compared with the standard method (std) for $\sqrt{\mathrm{s_{NN}}} =$ 7.7–19.6 GeV, as shown in Fig.~\ref{fig:scaledVar_2sub}. The $\eta$ gap between the two subevents is set to 0.4. The results, calculated for the particles within $|\eta|<$ 1.0, show that the values obtained using the two-subevent method are slightly suppressed compared to the standard method. However, in HIJING simulation for Pb+Pb collisions at LHC energies, the $k_{2}$ is suppressed by a factor of 3 when using the two-subevent method with particles selected from $|\eta|<$ 2.5~\cite{Bhatta:2021qfk}. This suppression is attributed to the decorrelation effect in the dynamic evolution of the fireball~\cite{Chatterjee:2017mhc}.

In high baryon chemical potential regions, baryons have similar statistics with mesons. To further explore the radial flow mechanism, we analyzed the scaled variance, skewness, and kurtosis at $\sqrt{\mathrm{s_{NN}}} =$ 19.6 GeV under three scenarios, ``hadron", ``baryon", and ``meson", as shown in Fig.~\ref{fig:scaledVarSkeKur_BM}. The results reveal that baryons exhibit more pronounced fluctuations in scaled variance, skewness, and kurtosis compared to mesons. This behavior might be attributed to the effects of radial flow, which preferentially accelerates heavier particles, pushing them from lower to higher $\pT$ values more effectively than lighter particles~\cite{Sarkar:2016ikv}.

\begin{figure}[htbp]
\centering
\includegraphics[scale=0.45]{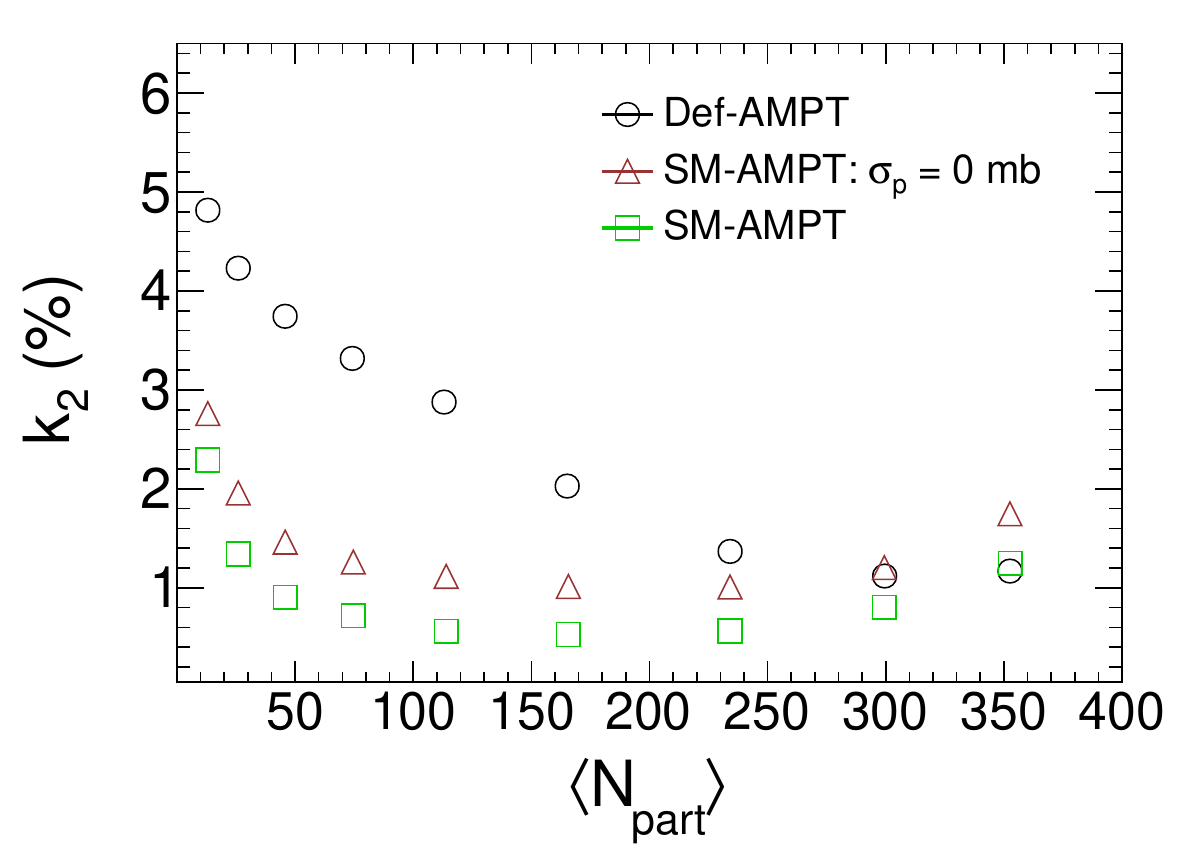}
\caption{(Color online) The scaled variance $k_2$ is calculated for Au+Au collisions as a function of $\mnpart$ at 19.6 GeV using three configurations: the standard method in default (Def-AMPT), normal string-melting (SM-AMPT), and string-melting without partonic interactions (SM-AMPT: $\sigma_p=$ 0 mb).}
\label{fig:scaledVarSkeKur_19.6}
\end{figure}

To dynamically investigate the factors contributing to EbE $\mpT$ fluctuations in the AMPT model, we compared the scaled variance $k_2$ in Au+Au collisions at $\sqrt{\mathrm{s_{NN}}} =$ 19.6 GeV using the standard method across three AMPT configurations: default version, the string-melting version without parton scattering (denoted as SM–AMPT: $\sigma_{p} = $0 mb), and the string-melting version with a parton cross section of 3 mb (SM–AMPT), as shown in Fig.~\ref{fig:scaledVarSkeKur_19.6}. 
The default configuration exhibits more pronounced $\mpT$ fluctuations compared to string-melting version, with and without partonic interactions. The discrepancy likely arises from differences in hadronization mechanisms. Furthermore, the minor variation between the string-melting AMPT versions with and without partonic interactions suggests that partonic evolutions slightly suppress the EbE fluctuations. 

\begin{figure}[htb]
\centering
\includegraphics[scale=0.45]{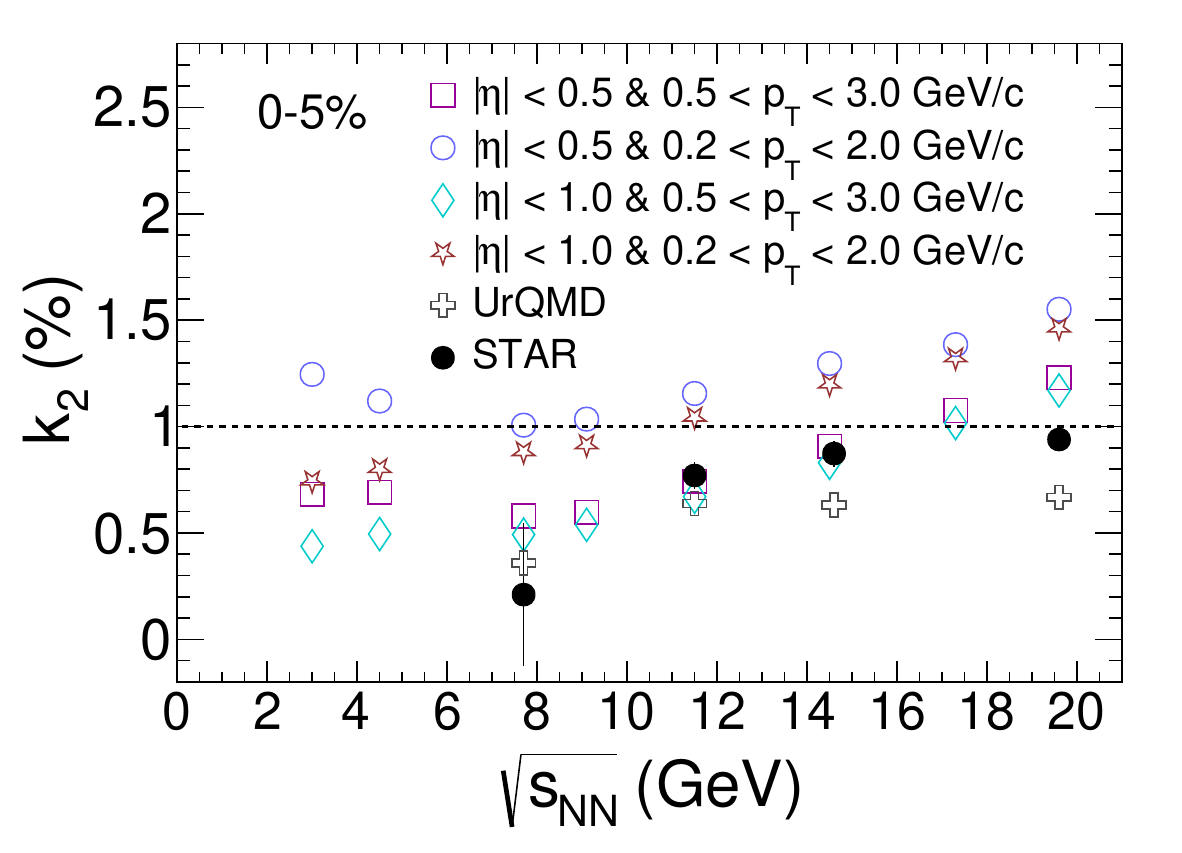}
\caption{(Color online) The scaled variance $k_2$ within different $\eta$ ranges, calculated using both standard and two-subevent methods for Au+Au collisions, is shown as a function of collision energy for the 0–5\% centrality. Results from AMPT simulations are compared to STAR experimental data~\cite{STAR:2019dow} and UrQMD calculations~\cite{STAR:2019dow}.}
\label{fig:VarScaled_E}
\end{figure}
Figure~\ref{fig:VarScaled_E} presents the scaled variance $k_2$ as a function of collision energy, ranging from 19.6 to 3.0 GeV in Au+Au collisions for the most central regions (0–5\%). Considering the varying transformation efficiency from initial geometry to momentum space across different $\pT$ ranges~\cite{Zhang:2017xda}, the analysis includes acceptance ranges including $|\eta|<$ 1.0 for 0.2 $<\pT<$ 2.0 GeV/c and 0.5 $<\pT<$ 3.0 GeV/c, as well as $|\eta|<$ 0.5 for the same $\pT$ intervals. 
For comparison, data from STAR experiment~\cite{STAR:2019dow} for $|\eta|<$ 0.5, along with results from UrQMD simulations are also included. Our findings show a significant energy dependence of scaled variances, with more pronounced suppression in the $|\eta|<$ 1.0 range compared to $|\eta|<$ 0.5. Furthermore, $k_2$ is more sensitive to the transverse momentum $\pT$ variations than to pseudorapidity $\eta$ acceptance. For 0.5 $<\pT<$ 3.0 GeV/c, the scaled variance in both $|\eta|<$ 1.0 and $|\eta|<$ 0.5 ranges are quantitatively consistent with STAR measurements.

Noted that as the incident energy decreases to $\sqrt{s_{\mathrm{NN}}}$ = 3.0 GeV in $|\eta|<$ 0.5, particularly within 0.2 $<\pT<$ 2.0 GeV/c, $k_2$ exhibits a significant abnormal increase compared to the values observed at $\sqrt{s_{\mathrm{NN}}}$ = 7.7 GeV. This indicates an enhancement of dynamical correlations at lower collision energies and highlights the potentiality to probe the properties of the medium formed during collisions.

\section{Discussions and Conclusion}
\label{sec:discussion}
The default and string-melting versions of the AMPT model have been improved by implementing a dynamic tuning mechanism for the Lund string fragmentation parameter $b_\mathrm{L}$, which now varies linearly with the impact parameter and exhibits increased sensitivity at higher beam energies. This improvement enables the AMPT model to accurately reproduce the trends in centrality and beam energy dependence of the mean transverse momentum $\mpT$ fluctuations. With this refined AMPT framework, we present a comprehensive systematic study of higher-order dynamical $\pT$ cumulants up to fourth order, using both standard and two-subevent methods as functions of $\mnpart$ and collision energies. 

Our analysis shows that $\pT$ cumulants up to fourth order, both normalized and unnormalized by $\dmpT$, exhibit a strong dependence on centrality across an energy range of $\sqrt{s_{\mathrm{NN}}}$ = 19.6 GeV to 3.0 GeV. Notably, the scaled variances as a function of centrality within $|\eta|<$ 0.5 align closely with STAR measurements. In most central collisions (0–5\%), the scaled variance is more sensitive to transverse momentum $\pT$ than to the acceptance $\eta$ range. Additionally, for the same $\pT$ range, the scaled variance demonstrates more significant suppression in the $|\eta|<$ 1.0 range compared to $|\eta|<$ 0.5. For collisions at $\sqrt{s_{\mathrm{NN}}}$ = 3 GeV, an enhancement in scaled variances is observed for 0.5 $<\pT<$ 3.0 GeV/c, particularly within $|\eta|<$ 0.5. This finding provides valuable references for the measurements from STAR fixed-target~\cite{RutikCPOD2024}, Beam Energy Scan program, and possible FAIR-CBM experiments.

We investigate the contributions to $\mpT$ fluctuations from baryons and mesons, separately. The results show that baryons exhibit considerable $\mpT$ fluctuations, likely due to their heavier masses, which make them more susceptible to the effect of radial flow compared to mesons. Additionally, we also evaluated two distinct hadronization mechanisms: Lund string fragmentation in the default version and quark coalescence in the string-melting version. The findings indicate that the default AMPT version demonstrates more pronounced $\mpT$ fluctuations. It would be interesting to measure in experiment the high-order transverse momentum fluctuations across collision energies.
 
\section{Acknowledgements} 
We would like to thank Ziwei Lin, Chao Zhang, and Liang Zheng for their insightful discussions. We thank Chen Zhong for the stimulating research environment. This work was supported in part by the National Key Research and Development Program of China under Contract Nos. 2022YFA1604900 and 2024YFA1612600, the National Natural Science Foundation of China (NSFC) under Contract Nos. 12205051, 12025501, 12147101, the Natural Science Foundation of Shanghai under Contract No. 23JC1400200, the Shanghai Pujiang Talents Program under Contract No. 24PJA009.
\bibliography{ref}{}
\bibliographystyle{apsrev4-1}

\end{document}